\documentclass{article}

\usepackage[preprint]{neurips}

\usepackage[utf8]{inputenc} % allow utf-8 input
\usepackage[T1]{fontenc}    % use 8-bit T1 fonts
\usepackage{hyperref}       % hyperlinks
\usepackage{url}            % simple URL typesetting
\usepackage{booktabs}       % professional-quality tables
\usepackage{amsfonts}       % blackboard math symbols
\usepackage{nicefrac}       % compact symbols for 1/2, etc.
\usepackage{microtype}      % microtypography
\usepackage{xcolor}         % colors
\usepackage{graphicx}

\AtBeginDocument{%
  }

\title{Supporting Creative Ownership through Deep Learning-Based Music Variation}

\author{%
  Stephen James Krol \\
  Monash University \\
  Melbourne, Australia \\
  \texttt{stephen.krol@monash.edu} \\
  \And
  Maria Teresa Llano \\
  University of Sussex \\
  Brighton, United Kingdom \\
  \And
  Jon McCormack \\
  Monash University \\
  Melbourne Australia \\
}

\begin{document}

\maketitle

\begin{abstract}
This paper investigates the importance of personal ownership in musical AI design, examining how practising musicians can maintain creative control over the compositional process. Through a four-week ecological evaluation, we examined how a music variation tool, reliant on the skill of musicians, functioned within a composition setting. Our findings demonstrate that the dependence of the tool on the musician's ability, to provide a strong initial musical input and to turn moments into complete musical ideas, promoted ownership of both the process and artefact. Qualitative interviews further revealed the importance of this personal ownership, highlighting tensions between technological capability and artistic identity. These findings provide insight into how musical AI can support rather than replace human creativity, highlighting the importance of designing tools that preserve the humanness of musical expression.
\end{abstract}

\section{Introduction}

The music industry is currently experiencing an \textsc{AI}-driven transformation, with various companies now providing complete song generation through prompts \citep{nayar2025ethics}. While the revolutionary potential of these technologies remains unclear, a more promising direction has emerged: using AI to support rather than replace human creativity \citep{otoole:2024,krol2025exploring}. These supportive applications, known as Artificial Intelligence Creativity Support Tools (\textsc{AI-CST}s), assist in the creative process, offering varying degrees of agency and control. Some tools are designed to automate routine or technically challenging tasks such as audio-to-midi transcription \citep{hawthorne2021sequence} and stem splitting \citep{spleeter2020}. These systems support the creative process but rarely provide creative ideas. Other tools instead offer more creative involvement, assisting in ideation and even the extension of musical ideas. Notable systems include the Multi-track Music Machine (\textsc{MMM}) \citep{ens2020flexible} which generates controllable multi-track MIDI; Music Transformer \citep{huang2018music}, offering melody harmonisation; and the Anticipatory Transformer \citep{thickstunanticipatory}, which supports infilling of musical phrases. 

Although some research has demonstrated the potential for AI-CSTs to support users \citep{louie2020novice}, recent work has highlighted the potential pitfalls of using AI for ideation. For example, \citet{Wadinambiarachchi2024} demonstrated how AI promotes \textit{design fixation}, ultimately limiting the conceptual space explored by the designer. Similarly, in their study on creative writing, \citep{Doshi2023} highlighted that while access to a Large-Language Model (LLM) improved the quality of writing for novices, it resulted in many homogenised outputs, reducing the overall diversity of the stories. \citet{mcguire2024establishing} expands on this and suggests that these limitations occur when the human becomes an \textit{editor} rather than an active \textit{co-creator}, highlighting the importance of humans actively directing the creative process. \citep{mcguire2024establishing} further argues that this active involvement can 
% "
``nurture intrinsic motivation'',
% ",
a vital component in human creativity \citep{amabile1985motivation}. 
Our earlier work supports this \citet{krol2025exploring}:
% . 
In a co-design study with musicians, we found that participants valued \textit{owning} the creative process and were unwilling to relinquish this control to a machine. Understanding how to design 
% Therefore, there is motivation to explore 
\textsc{AI-CST}s that promote active involvement in the creative process, by supporting, not replacing, creative skills, is therefore an important avenue of study.

In this work, we explore the themes of \textit{ownership} and \textit{active} co-creation in musical AI through a four-week ecological evaluation of a variation system, \textit{Rhapsody Refiner} \citep{krol2025exploring}. Unlike other symbolic systems \citep{ens2020flexible, huang2018music}, Rhapsody Refiner does not generate complete new ideas, instead, it relies on the musician to first provide a musical phrase, which it then explores. This forces the musician to be \textit{active} in both the initial ideation and final creation of the song. The findings from this study demonstrated the tool's usefulness was not in the generation of complete ideas, but instead \textit{moments} that sparked inspiration within the composer. Participants highlighted that this made them feel ownership of both the creative process and final composition, noting that this was important to their identities as composers. 

\begin{figure}[h]
  \centering
  \includegraphics[width=0.7\linewidth]{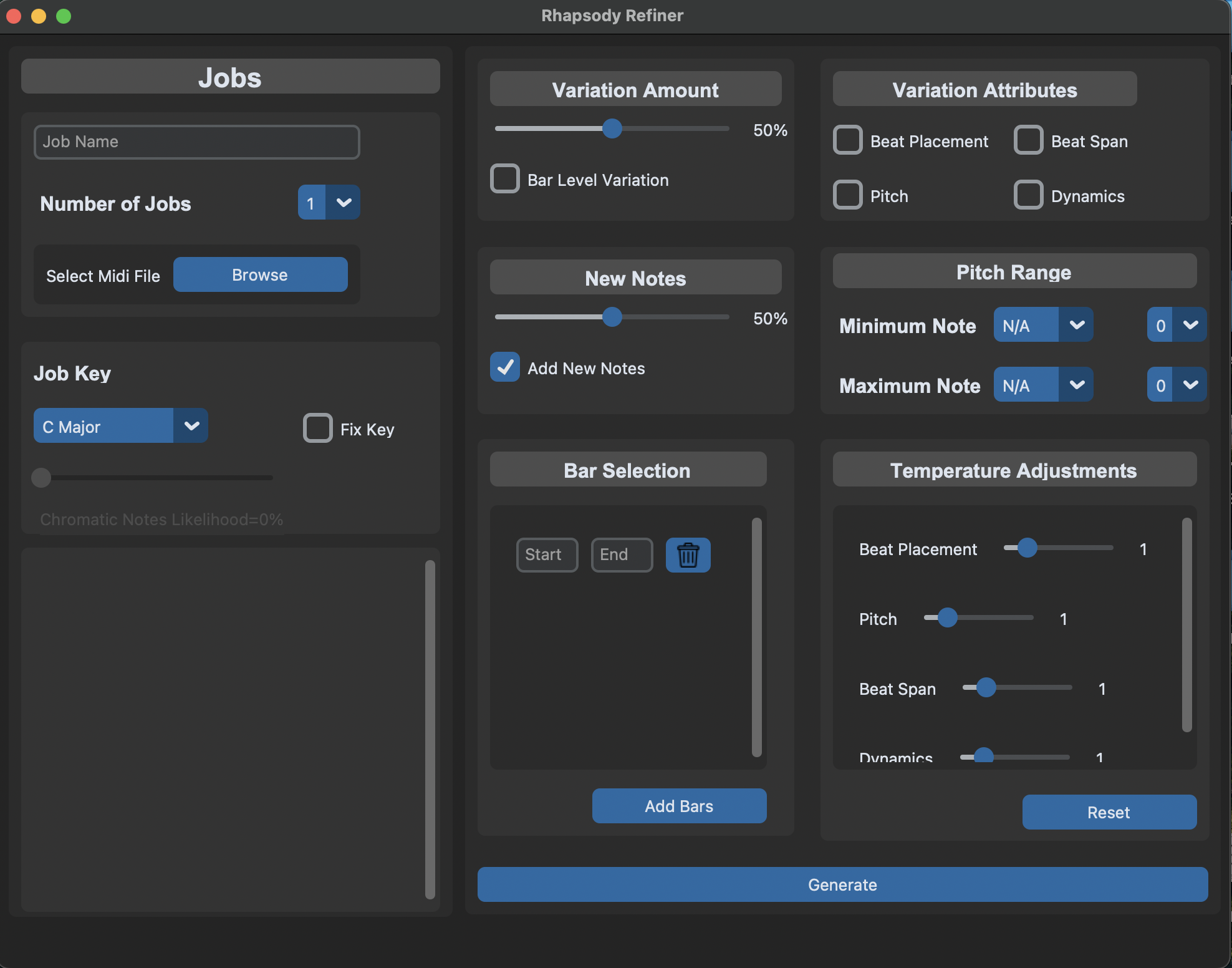}
  \caption{Interface for Rhapsody Refiner.}
  \label{fig:interface}
\end{figure}

\section{Rhapsody Refiner}
Rhapsody Refiner is a deep learning-based, music variation system that was co-designed with practising musicians \citep{krol2025exploring}. The tool was inspired by the MidiFormers project\footnote{https://github.com/tripathiarpan20/midiformers} and utilises MusicBert \citep{zeng2021musicbert}, a bidirectional transformer model trained for symbolic music understanding, to perform masked prediction on selected notes. The system supports  ideation by generating variations from user-provided phrases and includes functionalities for guiding the output. To generate a variation, users upload a MIDI file and set their desired parameters. The system uses these parameters to guide the masking of random sections of the song, which are then predicted using MusicBert. Participants can choose which aspects of a song to vary - for instance, altering the pitch of notes in a melody or adjusting the dynamics of the entire performance - and can generate multiple variations from the same input. Rhapsody Refiner was packaged as executable software that participants could run without any dependencies. 

At the time of evaluation, Rhapsody Refiner had been significantly improved from its previous version and was well suited for studying AI-assisted variation in composition, with co-design workshops providing early evidence of its usability \citep{krol2025exploring}. It is worth noting that for this system to be a complete tool for musicians, it would need to be implemented as a plugin for Digital Audio Workstations (DAW), however this was beyond the scope of this study. The interface for the system can be seen in figure \ref{fig:interface} and code for the backend is available on github\footnote{https://github.com/SensiLab/MusicVariationBert}.

% \subsection{Functionalities}
% Rhapsody Refiner offers various control functions to assist users in steering the variations. These functionalities were designed with practising musicians and are described in this section.

% \textbf{Variation Attributes}: Allows users to control which attributes of a note will be varied. In total there are four attributes, descriptions of these attributes are detailed in table \ref{tab:Attributes}.

% \textbf{Temperature Adjustments}: 

% \textbf{Variation Amount}: Controls the percentage of notes that will be masked and predicted by the system.

% \textbf{New Notes}: Allows the system to introduce new notes into the variation. New notes are added as a percentage of existing notes and spread equally over the available bars.

% \textbf{Bar Selection}: Controls which bar ranges the variation will be applied over. This ensures that users can provide the entire context of the song while still varying their ideal section. 

% \textbf{Pitch Range}: Restricts the system to predicting notes only within this specified pitch range. 

% \textbf{Key}: 

% \begin{table}[]
%     \caption{Note Attributes}
%     \label{tab:Attributes}
%     \centering
%     \begin{tabular}{ll}
%     \toprule
%       Attribute   &  Description \\
%       \midrule
%       Pitch & Controls the pitch of the note supporting 128 MIDI pitches \\
%       Beat Placement & Controls the onset of note in a bar with granularity of a 1/64 note \\
%       Beat Span & Controls the length of the note\\
%       Dynamics & Controls the velocity of the note\\
%       \bottomrule
%     \end{tabular}
% \end{table}

\section{Ecological Evaluation}
In order to understand how music variation could fit within composition, we conducted a four week ecological evaluation. Unlike a lab-based evaluation, ecological evaluations allow participants to experiment with software in their own creative environments for an extended period of time. This allows participants to develop a good understanding of the tool they are evaluating and has been used in other music-based studies \citep{habibi2022music,deruty2022development,krol2025exploring}.

In total, 8 practising musicians took part in the ecological evaluation. We defined a practising musician as a person that actively engages in the making of music, either professionally or as a hobbyist, and is distinguished from novice musicians who rarely engage in music making \citep{krol2025exploring}. Participants were recruited from known networks in the lab and were provided with a \$50 AUD voucher for their time. This study was also approved by the Monash University Human Research Ethics Committee (MUHREC).

% Participants were also asked to complete the General Sophistication section of the Goldsmiths Musical Sophistication Index (MSI) \citep{Mllensiefen:2014} to gain a basic insight into their musical understanding. An overview of participants can be seen in table \ref{tab:Participants}. 

Before commencing the study, participants were provided the software alongside a tutorial video that detailed the various functionalities of the system. Following their completion of the tutorial, participants were instructed to compose a song using Rhapsody Refiner, ensuring that they make consistent use of the system over the four-week period. Participants were recommended to constrain the song to a single instrument, though this was not required, and some participants chose to experiment with multitracked compositions. Participants were asked to keep a journal throughout the study to document key moments, and system log data was collected to provide additional insight into usage patterns. 

After completing the study, participants took part in a semi-structured interview that explored their usage of the tool. These interviews were transcribed and analysed using an inductive thematic analysis \citep{braun2006using} which involved two researchers independently coding the data and coming together to discuss the common themes present in the transcript. 

% In addition to this, participants were asked to complete a survey based on the work by \citet{tchemeube2023evaluating} which utilises various scales \citep{brooke1996sus,Cherry2014,davis1989perceived} to measure the perceived usability and acceptance of a creative tool.

\begin{table}[]
    \caption{Participant Musical Bacground}
    \label{tab:Participants}
    \centering
    \begin{tabular}{ll}
    \toprule
      Participant  & Perceived Musicality\\
      \midrule
      P1   & Songwriter and music producer \\
      P2 & Hobbyist with a background in classic piano \\
      P3  & Songwriter, drummer and member of a local band \\
      P4  & Electronic music producer \\
      P5  & Hobbyist songwriter and guitarist \\
      P6  & Composer and educator \\
      P7  & Jazz guitarist and member of local band \\
      P8  & Violinist and electronic instrument designer \\
      \bottomrule
    \end{tabular}
\end{table}

\section{Themes}

\subsection{A Tool for Moments}

Participants described Rhapsody Refiner as a \textit{tool for moments}. When interacting with the system, they found that 
% "
``it wasn't the whole eight bars of the variation [they] liked. It was just parts of them, that were really, really nice''
% "
(P1). These moments were described as 
% " 
``stunningly beautiful ideas... which in the context of the piece of music are huge''
% "
(P6). These ideas served as \textit{inspiration}, with participants noting that it was these 
% "
``small parts that helped inspire [them]''
% "
(P3) and that it took the composition 
% "
``in a couple of different, interesting ways that [they] probably wouldn't have done [themselves]''
% "
(P5). These moments were sometimes  found amongst what participants described as \textit{chaos}, which referenced phrases that felt 
% "
``really random''
% "
(P7) and 
% "
``quite scattered''
% " 
(P2). Participants noted that under some parameter settings this \textit{chaos} was overpowering, leading to moments that did not produce "many ideas" (P3). However, under the correct parameter settings, this \textit{chaos} could be effectively used to explore ideas, with P6 stating 
% "
``If we're going to play with this kind of thing, then it's actually important to have that randomness to come up with ideas that we wouldn't have come up with on our own. I think that's the whole bloody point, isn't it?''.
% ".

These imperfect outputs meant that participants needed to actively \textit{filter and refine} variations to fit their vision. One participant noted that at times 
% "
``[Rhapsody Refiner] had a good direction...it just didn't fully land the idea''
% "
and that it was up to the composer to 
% "
``help it get there''
% "
(P4). This meant that the system was \textit{reliant on the composer} to produce anything valuable. P7 noted that 
% "
``you have to create and be creative for it to work for you''
% " 
and that 
% "
``you can't put in garbage and expect to get anything better''.
% ".
This was echoed by other participants who noted that it could 
% "
``never give you a finished product''
% "
(P1) and that 
% "
``it is only as good as you are''
% "
(P3).

% *************************************

% Ultimately, participants found the tool to be \textit{useful}, particularly in \textit{ideation}: 
% % "
% ``when you're looking for a direction''
% % "
% (P4). P1 noted its value was in producing 
% % "
% ``sounds that I  really liked and wouldn't normally come up with''.
% % ". 
% This was echoed by other participants, such as P7, who mentioned that 
% % "
% ``part of the reason I was so happy with how good that voice leading was and how good those variations were is because when I generated them, I went, oh, I would not have thought of those myself''.
% % ".
% P7 added that a strength of the system is that it was also 
% % "
% ``showing me what I don't like'',
% % ", 
% helping them further refine their idea. Participants also identified situations where the system was \textit{not useful}, most notably when the composer already had a clear creative direction. P3 noted that when working on a song they had already written, they 
% % "
% ``were less open to new things''
% % "
% and that if they were to 
% % "
% ``use this in the future, it would be more for at the start of the songwriting process''.
% % ". 
% P7 also noted that it was never a solution to a problem as it 
% % "
% ``didn't ever really necessarily fix anything. It only added extra things''.
% % ".
% This again highlighted the system's reliance on the skill of the composer to produce anything of value.

% ********************************************

% talk about carsten inspired to change the time signature

\subsection{A Tool for Ownership}
A major theme explored throughout this study was the ownership of both the creative process and the artefact. When asked about how they felt regarding ownership, participants often noted that they felt \textit{control over the process} highlighting that they were 
% "
``the only one in control''
% " 
(P1) and that 
% "
``in terms of direction, yeah no, I feel like I was fully in control of that''
% " 
(P4). When asked why they felt this way participants referred back to the system being \textit{reliant on the composer}, with P4 stating that 
% "
``I feel like I've actually put effort in to bring a vision to life and I'm using Rhapsody to accentuate that vision and add variation to it''.
% ".
P7 expands on this, stating that 
% "
``I'm also sitting there, hearing what it's playing and saying, I do like this, I don't like this. I'm not sitting there being oh, the machine is telling me what to do... I don't feel like it's taken away from my creative process because I'm still there testing those variations''.
% ".

When asked about the ownership of the final artefact, there was less consensus amongst participants. Some participants felt they had \textit{complete ownership of the artefact}, such as P2 who stated ``Yeah, I think for this kind of system, I won't question the ownership because it's a variation system. I have to make something to variate. So everything was created from my own motif, which I composed myself''. This was echoed by P6 who stated that 
% "
``I'd be comfortable by saying, look, this is my work, supported. And some ideas emerged, but I still had to have the discernment to pick and choose those ideas and still shape it into a piece that is designed for human consumption''.
% ". 
Other participants instead felt that there was some level of \textit{joint ownership of the artefact}, such as P4 who attributed 10\% of the song to Rhapsody Refiner stating that 
% "
``I'd say it's a collaborative thing. Because it had this little effect''.
% ".
However, when asked why they attributed most of the song to themselves, P4 mentioned 
% "
``because it is not generating these ideas on its own in a vacuum. It's doing it based off something I've given it. And I think majority of the emotion, the feel of the track is coming from me''.
% ".
This concept of effort was also referred to by P3 who stated they felt more ownership over the song because while 
% "
``this change wouldn't have happened if Rhapsody didn't give me this idea, it was just an idea it gave me. I put it in all the work. So it sort of, it wouldn't exist without me or Rhapsody''.
% ".

% Participants also underlie a greater sense of ownership they felt working with Rhapsody in comparison to other state of the art tools; with P(Aidan) saying for instance that, with tools such as Suno, ``the AI will basically make it for you'' and that ``I'd feel genuinely bad trying to pass it off as my own work because it's like, I haven't actually put in any effort for the work.'', while with Rhapsody ``I'd probably be like, yeah, I wrote this and I used [Rhapsody] for some of the parts''. P(Al) also emphasized the open-ended nature of Rhapsody's controls, pointing out that they don't prescribe high-level musical outcomes, but instead offer low-level creative affordances: “It doesn't play with parameters like... the musical style you want? [in that case] we'd be starting to discuss a very different point.”

\subsection{A Tool for Creative Involvement}
% Participants indicated the feeling of \textit{retaining control} of the creative process when using Rhapsody. For instance, P(4) highlighted that with the system ``the artist gets the final say as to what goes into the music'' and contrasted it with the experience of using prompt-based systems ``you have control over [the direction], but you don't have control over the the minutia, like, you're not controlling the reverb, you're not designing the sounds''. This was echoed by other participants, with P(6) saying that ``I don't feel like it's stepping on my toes ... this tool doesn't dictate, it gives options'', while P(1) highlighted that ``the process is taking steps and making decisions, and I was the only one that did that.'' 

Being the driving force of the project was noticeably important to participants. When talking about alternative AI systems that had more creative agency, P6 noted that ``I find it uncomfortable because it takes away my sense of worth. So it's a very human experience, I fear that that music will lose novelty''. When asked why Rhapsody Refiner did not attack their musicianship P6 responded ``ultimately I'm the one shaping it. I think that's the key to it. That, my filtration mind identifies what works and what doesn't work, and then shapes that into the whole product''. This was echoed by other participants such as a P4 who highlighted that with Rhapsody Refiner `the artist gets the final say as to what goes into the music'' and contrasted it with the experience of using prompt-based systems where ``you have control over [the direction], but you don't have control over the the minutia, like, you're not controlling the reverb, you're not designing the sounds'' highlighting their desire to be part of the entire creative process.

When talking about songwriting, P7 noted that ``for me personally with songwriting, no song starts without your own idea. It will flash into your brain for half a second...it's like you hear it on the wind and you go, oh, what was that?...That idea, I want it to start with me''. The importance of creating this initial idea was referenced by other participants such as P3, who when asked if they would like a system that produced more ready-to-use variations, answered ``I don't think so...because then it feels a lot more like it's not all your song. Whereas what, I like, the most about using Rhapsody is it sort of pushes you into the right direction...so it still feels like yours... if it was just a full perfect [output] and I can just add that in my song, I wouldn't as much feel like that''.

\subsection{A Tool for Support} 

Ultimately, participants found the tool to be \textit{useful}, particularly in \textit{ideation}. On this, 
% participants noted the strengths of the system to help them start up the creative process: 
% % "
% ``when you're looking for a direction''
% % "
% (P4), 
P1 noted that part of its value was in producing 
% "
``sounds that I really liked and wouldn't normally come up with''.
% ". 
This was echoed by other participants, such as P7, who mentioned that 
% "
``part of the reason I was so happy with how good that voice leading was and how good those variations were is because when I generated them, I went, oh, I would not have thought of those myself’’.
% and P(Daniel) who stated that
% ``it took the melody in a couple of different, interesting ways and I wouldn't have probably done myself’’. 
Participants also praised the ability of the system to provide ideas on their {\em own} creative goals. On this, P5 highlighted that 
``the strengths come from when you have a specific thing you want input on, but you just don't know how to go about it’’, further emphasising that ``this is why it's good, because this is a tool for musicians''.
% , whereas some of those other tools are tools for people who want to make music without, you know, musical training or ability or whatever’’. 
% While P(Tace) stated that working with the system  
% ``felt like. I had a clear creative goal and I knew how to make it work [with Rhapsody Refiner]’’ % " and
P7 added that the system was also 
% "
``showing me what I don't like'',
% ", 
helping them further refine their idea. 

Despite these helpful features, the majority of participants did not believe that they could become over reliant on Rhapsody Refiner, praising this as an advantage of working with a variation system of this nature. For instance, when talking about relying on the system to compose music, P5 stated that 
``it's a sounding board more than anything’’ explaining that ``I don't see it being like an over-reliance thing because it doesn't give you like a complete song or anything. It gives you bits’’. P3 attributed this to being more like an ideation partner 
``I sort of see it as a replacement to having someone else there that's giving you ideas’’. 
Some of the limitations that participants highlighted also reinforce this idea, with P1 noting that 
``MIDI isn't a song [...] it's not complete. It's just like you have to do something with that MIDI file.’’, 
% and when participants identified situations where the system was \textit{not useful}, most notably when the composer already had a clear creative direction. For instance, 
while P3 noted that when working on a song they had already written, they 
% "
``were less open to new things''
% "
and that if they were to 
% "
``use this in the future, it would be more for at the start of the songwriting process''.
% ". 
P7 also noted that it was never a solution to a problem as it 
% "
``didn't ever really necessarily fix anything. It only added extra things''.
% ".
This again highlighted the system's reliance on the skill of the composer to produce anything of value.

\section{Discussion}
\subsection{Importance of Creative Ownership}
Throughout this study, it was clear that ownership, of both the process and artefact, was important to the musical identities of our participants. This concept has been supported in other studies \citep{tchemeube2023evaluating} and this ecological evaluation further demonstrates the importance of designing tools that support this ownership. Because Rhapsody Refiner relies on the user's creativity to be effective, participants often attributed ownership of the resulting song to themselves, emphasising the effort they invested in shaping the track. Participants noted that they would ``genuinely feel bad trying to pass [\textsc{AI} work] as [their] own'' (P4) highlighting that a strength of this system is ``that it doesn't totally take charge'' (P6). Participants also highlighted a reluctance to adopt tools that would create for them, noting they wanted to feel as though they were the main creative force behind the song.

Our study suggests that AI systems designed to support artistic ownership should require effort from their users. These systems should not lead but support, fitting within the creative process in a manner that does not automate, but instead suggests. Of course, not all musicians will care about ownership, and not every system needs to be designed with ownership in mind. However, for those who value a strong sense of authorship, designing to encourage effort may help preserve their creative agency.

% Participants noted a reluctance to adopt tools that would create for them, noting they wanted to feel as though they were the main creative force behind the song. In their paper \citet{norton2012ikea} introduced the \textsc{IKEA} effect, which described how consumers placed more value on goods they made themselves. Within 

\subsection{Imperfect Systems Encourage Active Co-Creation}

Our four-week ecological evaluation also demonstrated how a system like Rhapsody Refiner, which rarely produces perfect outputs, encouraged the user to be active throughout the co-creative process. The production of \textit{moments} rather than complete variations meant that participants had to use their own musical expertise to bring these ideas to life. In many cases, these ideas were not completely captured by the system but instead served as a stepping stone for a greater idea ``I feel like part of the process is something that happens outside of the variations'' (P5). The evaluation also highlighted the value of this interaction model, with participants emphasising the importance of randomness in their creative workflows. This stands in contrast to the prevailing paradigm in AI development, which prioritises the removal of imperfections in pursuit of consistent, ‘clean’ outputs \citep{ouyang2022training}.

Naturally, this reliance on the musician has its limitations. Some participants noted that if you have no musical skill or are a novice, it would be harder to extract use from Rhapsody Refiner. While some studies suggest that novices often gain the greatest benefit from AI \citep{Doshi2023}, the needs of novice and practising musicians are often different \citep{louie2020novice,becker2024designing,krol2025exploring} and thus require different design choices. Our study suggests that some practising musicians are more willing to embrace imperfect outputs if they lead to better ideas, drawing on their skills to refine and implement those ideas independently. This is in contrast to novice musicians who require more predictability \citep{louie2020novice} and who have less ability to turn moments into finished songs.

% talk specifically about not needing perfect outputs

\subsection{Maintaining Humanity in Music Composition}

The making and consuming of music is a unique human practice. It plays an important role in our society as a means for self-expression \citep{denora1999music}, social interaction \citep{Rabinowitch2021} and therapy \citep{lee2016effects}. Musicians take pride in making music and despite what some leaders in the technology sector might say\footnote{SunoAI CEO on a recent podcast - "I think the majority of people don’t enjoy the majority of time they spend making music"} - enjoy the process of making music \citep{hill2016until}. When asked about how P6 felt regarding the emergence of \textsc{AI} in their practice, they responded ``Look, I don't like it from a selfish point of view and also human egocentric point of view, I'd like to think that as a species we've got something that's a little bit more unique. This really challenges that sense of what does it mean to be human?''. This response reflects not only a resistance to adopt \textsc{AI} in music composition, but also deeper anxieties regarding their identity. Understandably, many would hesitate to embrace a technology that threatens to render their own role obsolete, creative or not. However, these anxieties go beyond self preservation and instead question why we would want to automate something that is so innately human. 

In our study, these anxieties where not extended to Rhapsody Refiner. P6 noted "Your program, I really like that it can support exploring different ideas. That there is a certain randomness around the factoring of it". The tool’s reliance on the composer meant it could only play a supportive role, leaving creative control firmly in the user's hands. As our study confirmed, this was partly because Rhapsody Refiner does not generate an initial idea, instead assigning that responsibility to the user. By not taking this away from them, participants felt more comfortable using the system, feeling that the essence of the music came from them. It's worth noting that to our participants, the idea was more than just a prompt. To them, it represented music that already existed in their minds, a creative vision that they brought to the system, rather than just a lingual blueprint for generating music. 

% In a time where the overuse of AI in creative contexts has been shown to erode cognitive skills like divergent thinking \citep{microsoftStudy:2025, lixiang:2025} and undermine individuals’ confidence in their own creative abilities \citep{tang2024best, stillcreative:2025}, it is important that we support the design of technology that supports creative practice, while still preserving a sense of personal identity within creatives.

% P7 pointed out that "anyone can make an arrangement and just go through every single possible variation, it's more time consuming..."

% ``I think this is why it's good because this is a tool for musicians'' (P5)

% While the results from this study cannot be generalised to all musicians, other studies \citep{krol2025exploring} demonstrate that there are groups of musicians that have no

% P6 emphasised the need for distinct roles between humans and AI in composition, stating, ``I do think there's a real separation, the emotional core of a human is really our unique feature'', and highlighting that AI should take a more backseat role.

% Continuing on this thread, P6 also noted ``I do think there's a real separation, the emotional core of a human is really our unique feature''. 

% a role for AI and human, humans should be in charge of the emotion

% Yeah, I do think there's a real separation. The emotional core of a human is really our unique feature

\section{Limitations and Future Work}
Due to unexpected funding constraints, participant compensation was lower than originally planned. This made it hard to recruit participants for the ecological evaluation and limited our ability to run quantitive evaluations. Future work will focus on this and examine whether a system like Rhapsody Refiner enables composers to produce higher-quality music, as judged by experts, compared to systems that provide more automation capabilities.

\section{Conclusion}
Our findings highlight that tools like Rhapsody Refiner, which preserve creative ownership, encourage active co-creation, and maintain a clear human role in music-making, can foster meaningful and satisfying artistic engagement with music AI systems.  By requiring effort, embracing imperfections, and supporting rather than replacing, Rhapsody Refiner helped preserve a sense of personal identity within our participants while still providing use in their practice. 

\bibliographystyle{ACM-Reference-Format}
\bibliography{neurips.bib}

\end{document}